\documentclass{article}

 \usepackage[preprint, nonatbib]{nips_2018}

\usepackage[utf8]{inputenc} 
\usepackage[T1]{fontenc}    
\usepackage{hyperref}       
\usepackage{url}            
\usepackage{booktabs}       
\usepackage{amsfonts}       
\usepackage{amsmath}
\usepackage{color}
\usepackage{nicefrac}       
\usepackage{microtype}      
\usepackage[pdftex]{graphicx}

\newcommand{\dif}{\textup{d}}
\newcommand{\me}{\textup{e}}

\newcommand{\xx}{{\mathbf{x}}}

\newcommand{\zz}{{\mathbf{z}}}

\newcommand{\RR}{\mathbb{R}}

\newcommand{\figref}[1]{Figure \ref{#1}}

\title{A Nonstationary Designer Space-Time Kernel}

\author{
  Michael McCourt \\
  SigOpt\\
  San Francisco, CA, USA \\
  \texttt{mccourt@sigopt.com} \\
  \And
  Gregory Fasshauer, David Kozak \\
  Colorado School of Mines \\
  Golden, CO, USA \\
  \texttt{\{fasshauer, dkozak\}@mines.edu} \\
}

\begin{document}

\maketitle

\begin{abstract}
    In spatial statistics, kriging models are often designed using
    a stationary covariance structure; this translation-invariance
    produces models which have numerous favorable properties.  This
    assumption can be limiting, though, in circumstances where the
    dynamics of the model have a fundamental asymmetry, such as in
    modeling phenomena that evolve over time from a fixed initial
    profile.  We propose a new nonstationary kernel which is only
    defined over the half-line to incorporate time more naturally
    in the modeling process.
\end{abstract}

\section{Introduction}

Covariance kernels for Gaussian random fields can be defined in
a number of ways---see, e.g., \cite{RasmussenWilliams06} or
\cite{FasshauerMccourt2015}.
In spatial statistics applications in $\RR^d$,
kernels are often \emph{radial}: the covariance
between random field values at two points $\xx,\zz\in\RR^d$
is defined through a sense of distance between $\xx$ and $\zz$
\cite{stein2012interpolation}.

Radial covariance kernels (sometimes referred to as radial
basis functions) have numerous
advantageous properties, such as computational efficiency and
Bochner's theorem of positive definiteness
\cite{fasshauer2007meshfree, scholkopf2002learning}.
However, they also yield stationary kriging
models, which is not always ideal.  Some work has been done
to try to build nonstationary models from stationary kernels
(see, e.g., \cite{martinez2017bayesian} or \cite{le2017zooming}).

Another strategy to permit nonstationary models is to
use nonstationary kernels, e.g., polynomial kernels
\cite{shawe2004kernel}.
Strategies such as power series
kernels \cite{zwicknagl2009power}, Green's kernels
\cite{fasshauer2013reproducing},
and kernels derived from
weighted Sobolev spaces \cite{dick2014lattice} can frequently
(though not necessarily) yield nonstationary kernels.  Additionally,
complicated (possibly nonstationary) kernels can be constructed
from simpler (possibly stationary) kernels through addition
or multiplication (see \cite{berlinet2004reproducing} for a
thorough discussion and \cite{malkomes2016bayesian} for
recent applications) or through transforms \cite{remes2017non}.

One of the benefits of designing special kernels through
strategies such as weighted/generalized Sobolev spaces is to
create kernels which satisfy certain properties such as
smoothness, locality or boundary behavior.  In this paper, we
adapt a strategy for designing kernels in a bounded domain to
design a kernel on the half-line.  We explore the properties
of this kernel as well as its viability for contributing to
modeling space-time phenomena.

\section{Mercer series kernels\label{sec:mercersserieskernels}}
Given a compact domain $\Omega\subseteq\RR^d$,
Mercer's theorem says that all positive definite kernels
(i.e., covariance kernels) $K:\Omega\times\Omega\to\RR$
have the series representation
\begin{align}
    \label{eq:mercerseries}
    K(\xx,\zz) = \sum_{n=1}^\infty \lambda_n
        \varphi_n(\xx)\varphi_n(\zz),
\end{align}
where $\lambda_n, \varphi_n$ are the eigenvalues and eigenfunctions
of the integral eigenvalue problem with the associated
orthonormality condition
\begin{align}
    \label{eq:integraleigenvalueproblem}
    \int_\Omega K(\xx, \zz)\varphi_n(\xx)\rho(\xx)\,\dif\xx =
        \lambda_n\varphi_n(\zz), \quad
    \int_\Omega \varphi_m(\xx)\varphi_n(\xx)\rho(\xx)\,\dif\xx =
        \delta_{m,n},
\end{align}
for a weight function $\rho:\Omega\to\RR_+$ such that
$\int_\Omega\rho(\xx)\,\dif\xx=1$.
If $\lambda_n>0$, $K$ will be positive definite.

This kernel decomposition is often used theoretically to invoke
the ``kernel trick'' for support vector machines
\cite{scholkopf2002learning}.  In \cite{RasmussenWilliams06},
eigenfunctions for the popular squared exponential (Gaussian)
kernel were discussed.  Later, in
\cite{cavoretto2015introduction} and
\cite[Chapter 3.9]{FasshauerMccourt2015},
new nonstationary kernels were designed using
eigenfunctions defined through a differential equation and
desired orthogonality properties, respectively.

\section{A kernel defined on the half-line\label{sec:newkernel}}

In this section, we design a kernel for which
$\Omega=[0,\infty)$; this will
match the structure of most modeling problems in time, where there
is a distinct starting time for observing data, but no distinct
end.  Note the use of $t$ in place of $\xx$ because we will primarily apply this kernel to problems in time.

\subsection{Eigenfunctions from the Laguerre polynomials
\label{sec:laguerreeigenfunctions}}
We revisit the concept from \cite[Chapter 3.9]{FasshauerMccourt2015},
but we replace the Chebyshev with the Laguerre polynomials and
their associated orthogonality condition (for $\alpha>-1$)
\begin{align}
    \label{eq:laguerrepolynomials}
    L_n (t) = \sum_{i=0}^n (-1)^i
    \genfrac{(}{)}{0pt}{0}{n+\alpha}{n-i}
    \frac{t^i}{i!},
    \quad
    \int_0^\infty t^\alpha \me^{-t} L_m(t)L_n(t)\,\dif t = 
    \frac{\Gamma(n+\alpha+1)}{\Gamma(n+1)} \delta_{m,n}.
\end{align}

Analogously to the Gaussian eigenfunctions from
\cite{fasshauer2012stable}, we choose the
eigenfunctions to have the form
\[
    \varphi_n(t) = \gamma_n\me^{-\delta t}L_n(t),
\]
for $0<\delta<\frac{1}{2}$ and a normalization constant
$\gamma_n$.  We seek to enforce the orthogonality in
\eqref{eq:integraleigenvalueproblem} utilizing the
Laguerre polynomial orthogonality identity
in \eqref{eq:laguerrepolynomials}; this suggests the
weight function
\[
    \rho(t) = t^\alpha\me^{-(1-2\delta)t}
        \frac{(1-2\delta)^{\alpha+1}}{\Gamma(\alpha+1)},
\]
which satisfies $\int_0^\infty\rho(t)\,\dif t=1$.  Using
this we can enforce orthonormality to identify $\gamma_n$,
\[
    \int_0^\infty \varphi_n(t)^2\rho(t)\,\dif t =
        1
    \quad\Rightarrow\quad
    \gamma_n = \sqrt{
        \frac{\Gamma(n+1)}{\Gamma(n+\alpha+1)}
        \frac{\Gamma(\alpha+1)}{(1-2\delta)^{\alpha+1}}
    }.
\]

Laguerre polynomials are a natural choice for the eigenfunctions
because they are orthogonal on the domain $[0,\infty)$.  This yields a kernel
which is inherently defined only on $[0,\infty)$ which is
a more logical choice than simply restricting a kernel
defined on $\RR$ (such as the Mat\'ern) into $[0,\infty)$.
Other functions which form an orthonormal basis over
$[0, \infty)$, such as the Bessel functions $J_k$, could also
potentially be used to create different $\varphi_n$.

\subsection{Forming a kernel from the eigenfunctions
\label{sec:formingthekernel}}

The Mercer series \eqref{eq:mercerseries} requires both
eigenfunctions $\varphi_n$ and eigenvalues $\lambda_n$.
As was the case in
\cite[Chapter 3.9]{FasshauerMccourt2015}, we have freedom in
how we choose the eigenvalues; these choices will affect
the smoothness of the resulting kernel.  Here, we choose
the eigenvalues
\[
    \lambda_n = (1-\omega)\,\omega^n, \qquad 0<\omega<1,
\]
which satisfies the positive definite criteria that
$\lambda_n>0$ and $\sum_{n=0}^\infty\lambda_n=1<\infty$.
Because these eigenvalues decay geometrically, we expect
$K(\cdot, s)\in C^\infty(\Omega)$ \cite{Reade84}.

At this point, the kernel is fully specified
by \eqref{eq:mercerseries}; however, it is possible to define
a closed form for these specific eigenvalues, so that the
infinite series need not be approximated.  We can utilize
the generating function of the Laguerre polynomials to write
the Hardy--Hille formula \cite{watson1933notes}
\begin{align}
    \label{eq:hardyhille}
    \sum _{n=0}^{\infty }
    \omega^{n}
    \frac {\Gamma(n+1)}{\Gamma \left(1+\alpha +n\right)}\,
    L_{n}(t)L_{n}(s)=
    \frac {1}{\left(ts\omega\right)^{\frac {\alpha }{2}}
    (1-\omega)}
    \me^{-\frac {\left(t+s\right)\omega}{1-\omega}}
    I_\alpha\left(\frac {2{\sqrt {ts\omega}}}{1-\omega}\right),
\end{align}
where $I_\alpha$ is a modified Bessel function of the first
kind \cite[Chapter 10.25]{NIST_DLMF}.
The Mercer series is
\[
    K(t, s) = \sum_{n=0}^\infty \lambda_n\varphi_n(t)\varphi_n(s) =
        (1-\omega)
        \frac {\Gamma(\alpha+1)}{(1-2\delta)^{\alpha+1}}
        \me^{-\delta(t+s)}
        \sum_{n=0}^\infty \omega^n
        \frac {\Gamma(n+1)}{\Gamma \left(1+\alpha +n\right)}
        L_n(t)L_n(s)
\]
so substituting in \eqref{eq:hardyhille} gives the kernel in
its closed form
\begin{align}
    \label{eq:closedformkernel}
    K(t, s) = 
        \frac{\Gamma(\alpha+1)}{(1-2\delta)^{\alpha+1}}
        \left(ts\omega\right)^{-\alpha/2}
        \me^{-(t+s)\left(\delta+\frac{\omega}{1-\omega}\right)}
        I_{\alpha }\left({\frac {2{\sqrt {ts\omega}}}{1-\omega}}\right).
\end{align}

\subsection{Computing the kernel safely\label{sec:computing}}
The computation of the kernel in \eqref{eq:closedformkernel}
may be complicated by  indeterminate forms
and competing terms of exponential magnitude.
The latter complication can be alleviated by
computing $\exp(\log(K(t, s)))$;
$\log(K(t, s))$ can be safely computed by
using Stirling's approximation
\cite[Chapter 5.11.1]{NIST_DLMF} to compute
$\log(\Gamma(\alpha + 1))$ and leveraging the
decomposition $I_\alpha(u)=\me^u\tilde{I}_\alpha(u)$
where $\tilde{I}_\alpha(u)$ is bounded and can be computed
safely \cite{amos1986algorithm}. 

We can resolve indeterminate forms by
utilizing the asymptotic forms of $I_\alpha$,
\[
    I_\alpha(u) \sim
        \frac{1}{\Gamma(\alpha + 1)}
        \left(\frac{u}{2}\right)^\alpha
        \quad\text{as}\;u\to0,\qquad\qquad
    I_\alpha(u) \sim \frac{\me^u}{\sqrt{2\pi u}}
        \quad\text{as}\;u\to\infty,
\]
from \cite[Chapter 10.30.1]{NIST_DLMF}
and \cite[Chapter 10.40.1]{NIST_DLMF}, respectively.
This gives the behavior of $K(\cdot,s)$ for a fixed
$s\in[0, \infty)$ as
\begin{subequations}
\begin{align}
    \label{eq:kernellimitzero}
    K(t, s) &\sim 
        \frac{1}{(1-2\delta)^{\alpha+1}(1-\omega)}
        \me^{-(t+s)\left(\delta+\frac{\omega}{1-\omega}\right)}
        \quad\text{as}\;t\to0, \\
    K(t, s) &\sim
        \frac{\Gamma(\alpha+1)}{2(1-2\delta)^{\alpha+1}}
        \sqrt{\frac{1-\omega}{\pi(ts\omega)^{1/2+\alpha}}}
        \me^{-\left(\delta+\frac{\omega}{1-\omega}\right)
            (\sqrt{t}-\sqrt{s})^2
            -2\left(\delta-
                \frac{\sqrt{\omega}}{1+\sqrt{\omega}}
                \right)
            \sqrt{ts}
            }
        \quad\text{as}\;t\to\infty.
    \label{eq:kernellimitinfinity}
\end{align}
\end{subequations}

\subsection{Free parameters and their
    interpretations/implications\label{sec:freeparameters}}
    
As is the case with all positive definite kernels, this
new kernel has free parameters which characterize its
behavior: $\alpha>-1$, $0<\delta<1/2$, and $0<\omega<1$.
The interplay
between these parameters is nontrivial.  We state here
some initial analyses which we hope lay the foundation
for future work.

\begin{figure}[ht]
    \centering
    \includegraphics[width=.3\linewidth]{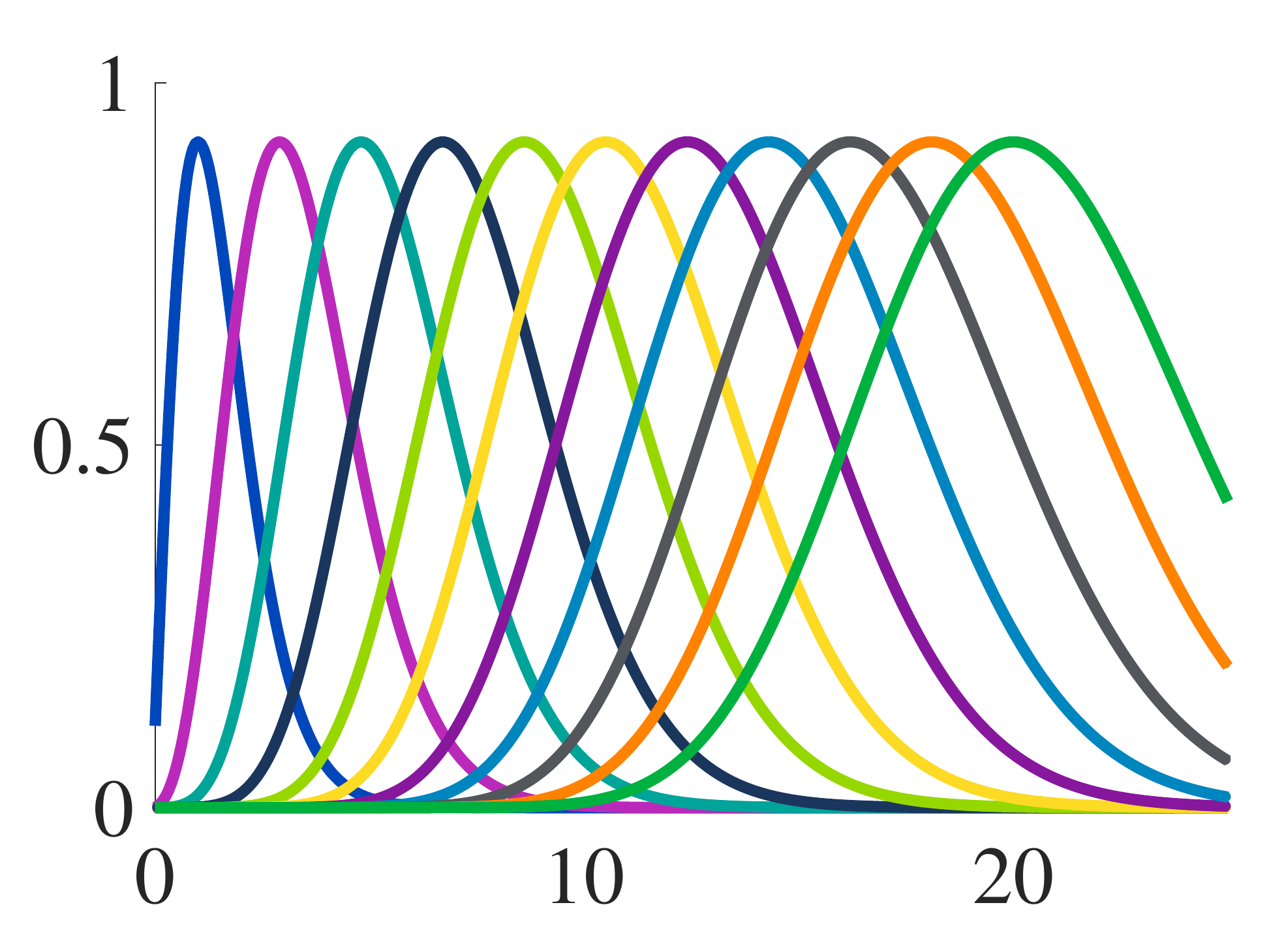}
    \includegraphics[width=.3\linewidth]{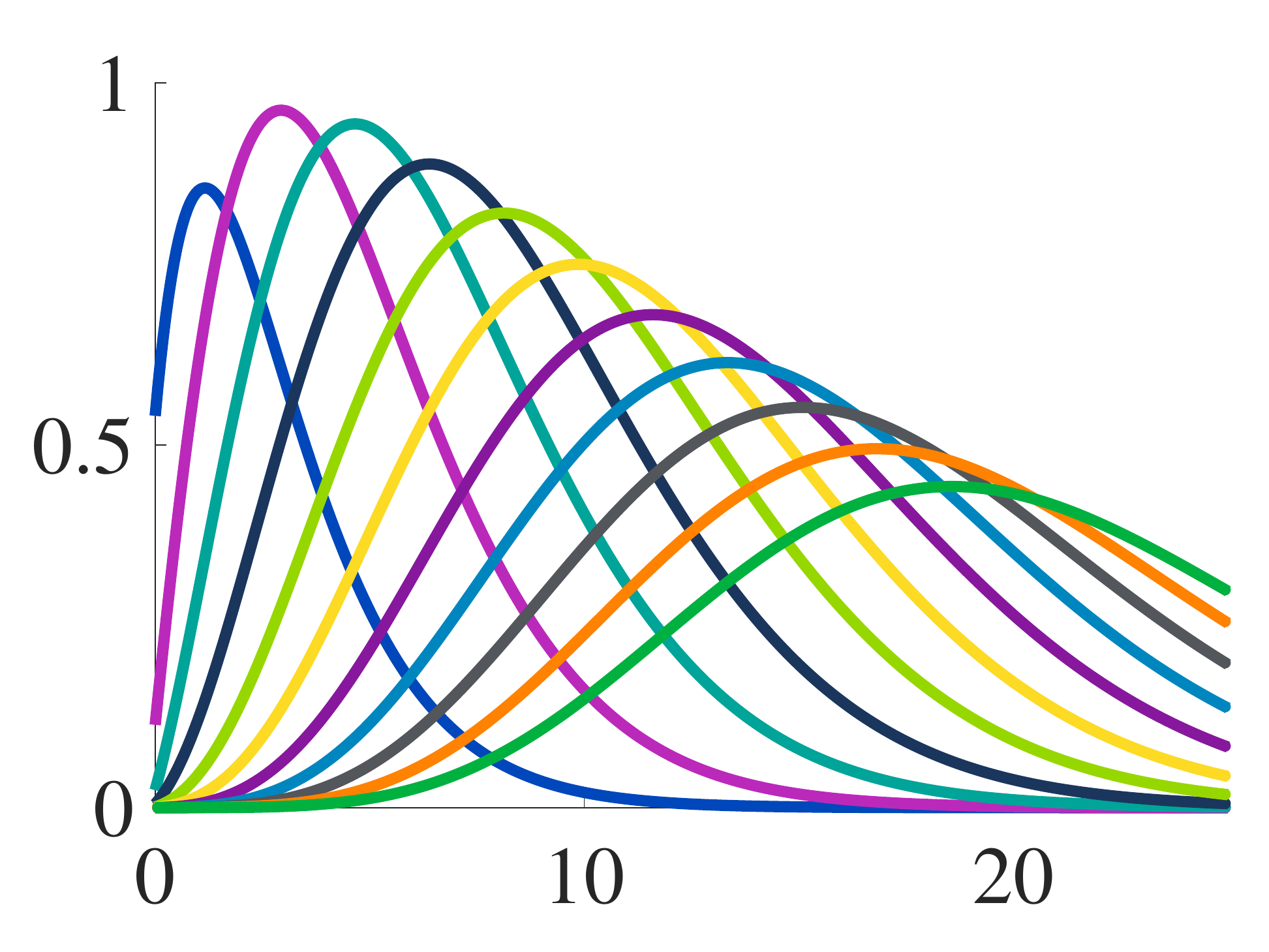}
    \includegraphics[width=.3\linewidth]{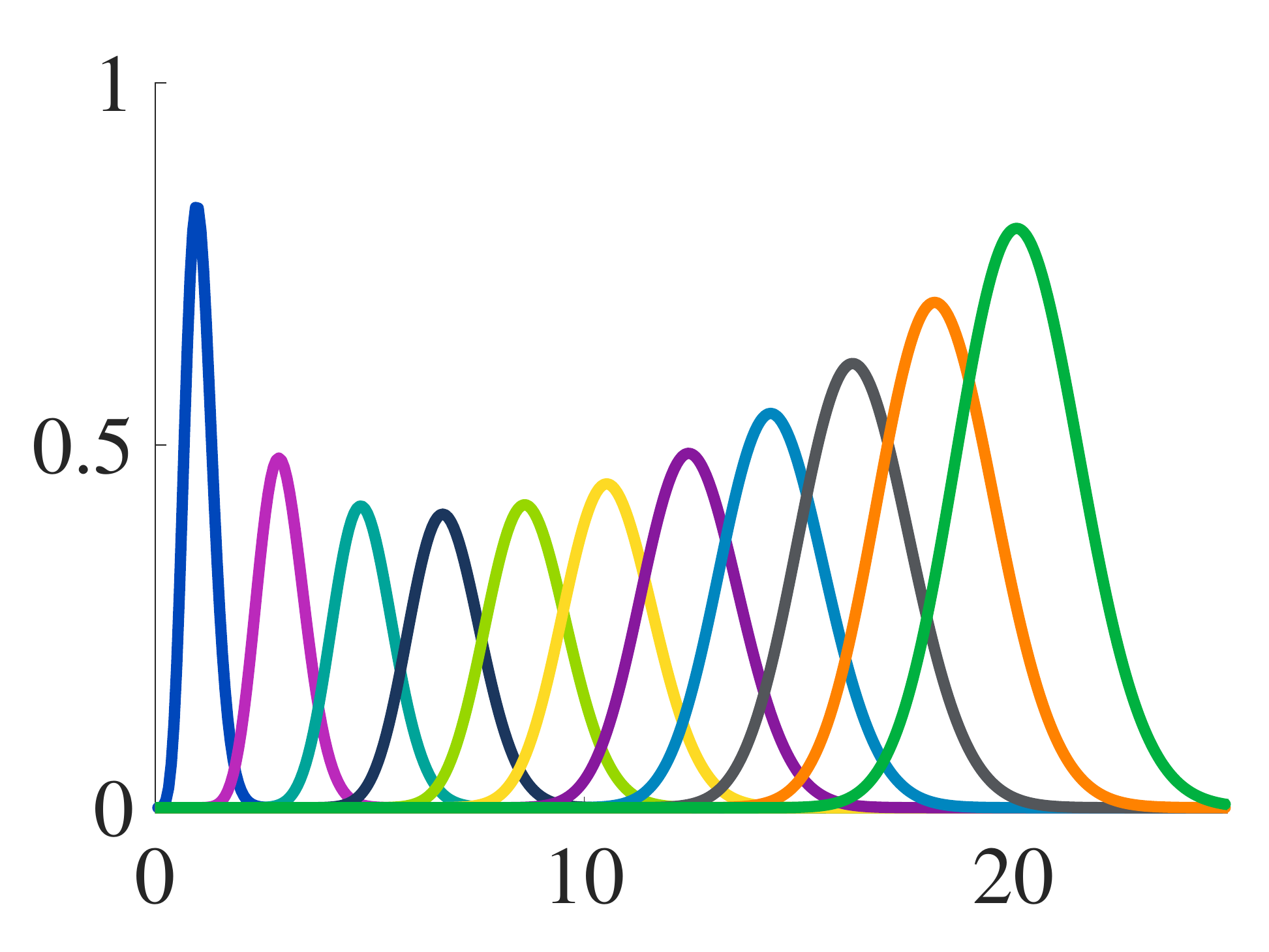}
    \caption{
        \emph{left}: $\alpha=-.5$, $\delta=.455$, $\omega=.7$;
        \emph{center}: $\alpha=-.7$, $\delta=.389$, $\omega=.3$;
        \emph{right}: $\alpha=.2$, $\delta=.439$, $\omega=.95$.
        \label{fig:kerneldemo}
    }
\end{figure}

The behavior  is easiest to analyze when
the argument of $I_\alpha$ is large and
\eqref{eq:kernellimitinfinity} is appropriate.  This can be
true for both $t\to\infty$ or for any $t$ with large enough
$s$, i.e., when observed data is sufficiently far from
the origin.  Some insights follow below
(encapsulated somewhat in \figref{fig:kerneldemo}).
\begin{itemize}
    \item The exponential term guides the long-distance
        behavior.  Its impact varies based
        on the relationship between
        $\delta$ and $\sqrt{\omega}/(1+\sqrt{\omega})$.
        Regardless of the value of $\delta$,
        the first term in the exponential dominates
        the second for any fixed $s$ as $t\to\infty$.
        implying that the kernel always goes to 0 at
        $\infty$.
    \item Similarly, the behavior of the $ts\omega$ component
        depends on $\alpha$ being greater than/less than/equal to
        -1/2.  For $\delta\neq\sqrt{\omega}/(1+\sqrt{\omega})$,
        the $t\to\infty$ behavior of the kernel is dominated
        by the exponential term and $\alpha$ plays only
        a scaling role.
    \item Setting $\alpha=-1/2$ and
        $\delta=\sqrt{\omega}/(1+\sqrt{\omega})$ yields a kernel
        which has only a $\sqrt{t}-\sqrt{s}$ contribution; this
        implies that
        $\lim_{s\to\infty}\max_{t\geq0}K(t, s)=(1+\sqrt{\omega})/2$.
    \item The $\omega$ term acts as a sort of inverse length scale
        (or shape parameter) for the kernel.
        The interpretation of it as a length scale is
        complicated for $s$ near the origin, but more obvious
        as $s$ grows (for observed results far from the origin).
\end{itemize}


\section{Example: Modeling west coast temperature data\label{sec:example}}
The kernel in \eqref{eq:closedformkernel} must be
supplemented by components meant to model spatial
phenomena.  One natural mechanism for facilitating
this is with a product kernel, consisting of separate
time and space components (discussed in
\cite{CressieHuang99} and \cite{gneiting2006geostatistical}).
Here, we consider a tensor product
between a Gaussian $K_g$ in space
and the
kernel \eqref{eq:closedformkernel}, denoted here
as $K_\ell$, in time:
$
K((\xx\;t),\,(\zz\;s))
 =
K_g(\xx,\zz)\, K_\ell(t,s).
$

Our example comes from the GEFS Reforecast dataset \cite{NOAA};
in particular, we use daily measurements for eight days of
surface temperature data across the western United States
with $1^{\circ}\times 1^{\circ}$ resolution.
After training on seven days, we
compute the Gaussian random field posterior mean
on the eighth day.
We use this as a prediction at each
of the $28\times29$ grid points and assess the root mean
squared error (RMSE) of that prediction.
This amounts to 5684 training points and 812 test points.

\begin{figure}[ht]
    \centering
    \includegraphics[width=.35\linewidth]{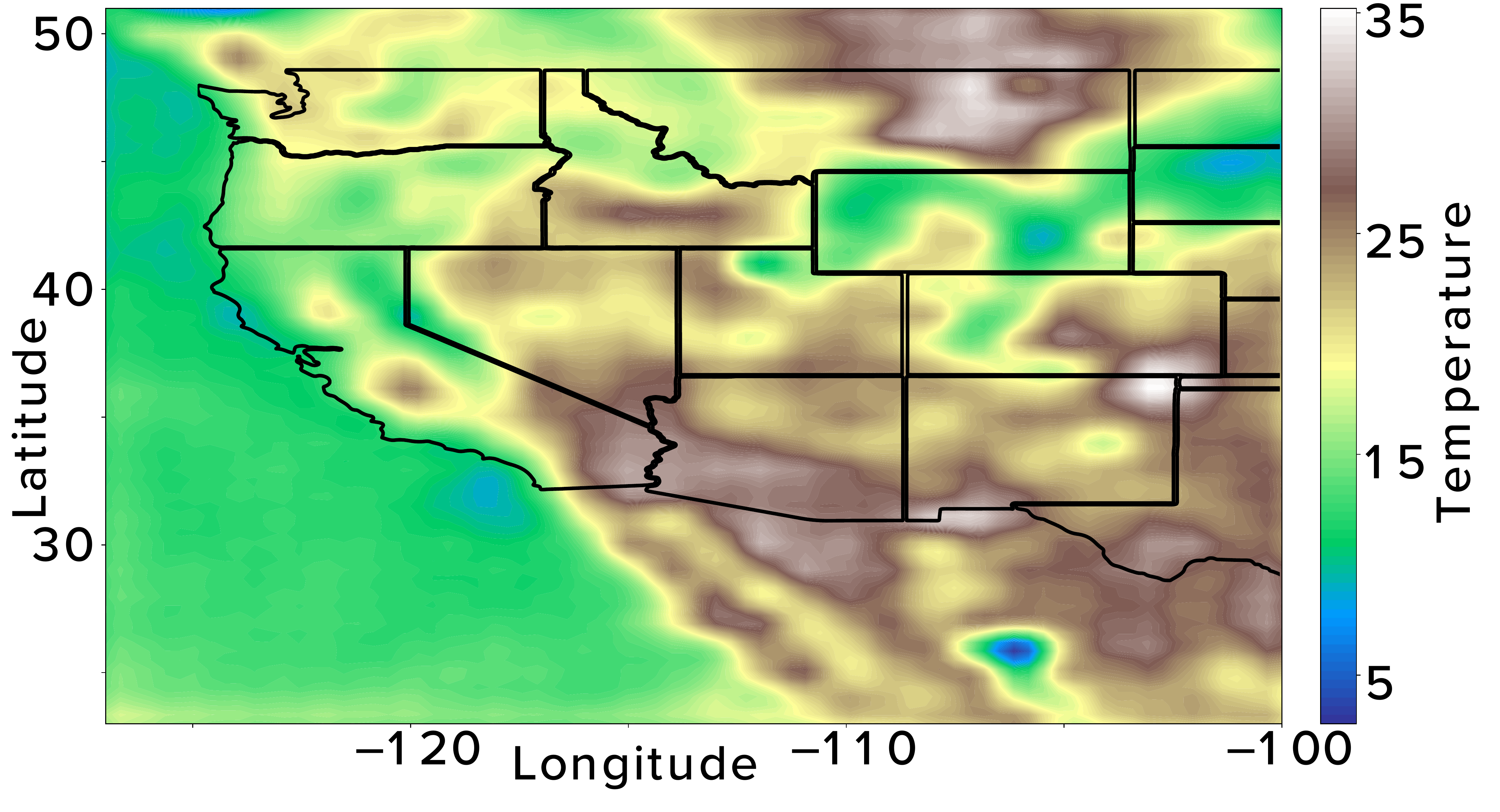}
    \includegraphics[width=.3\linewidth]{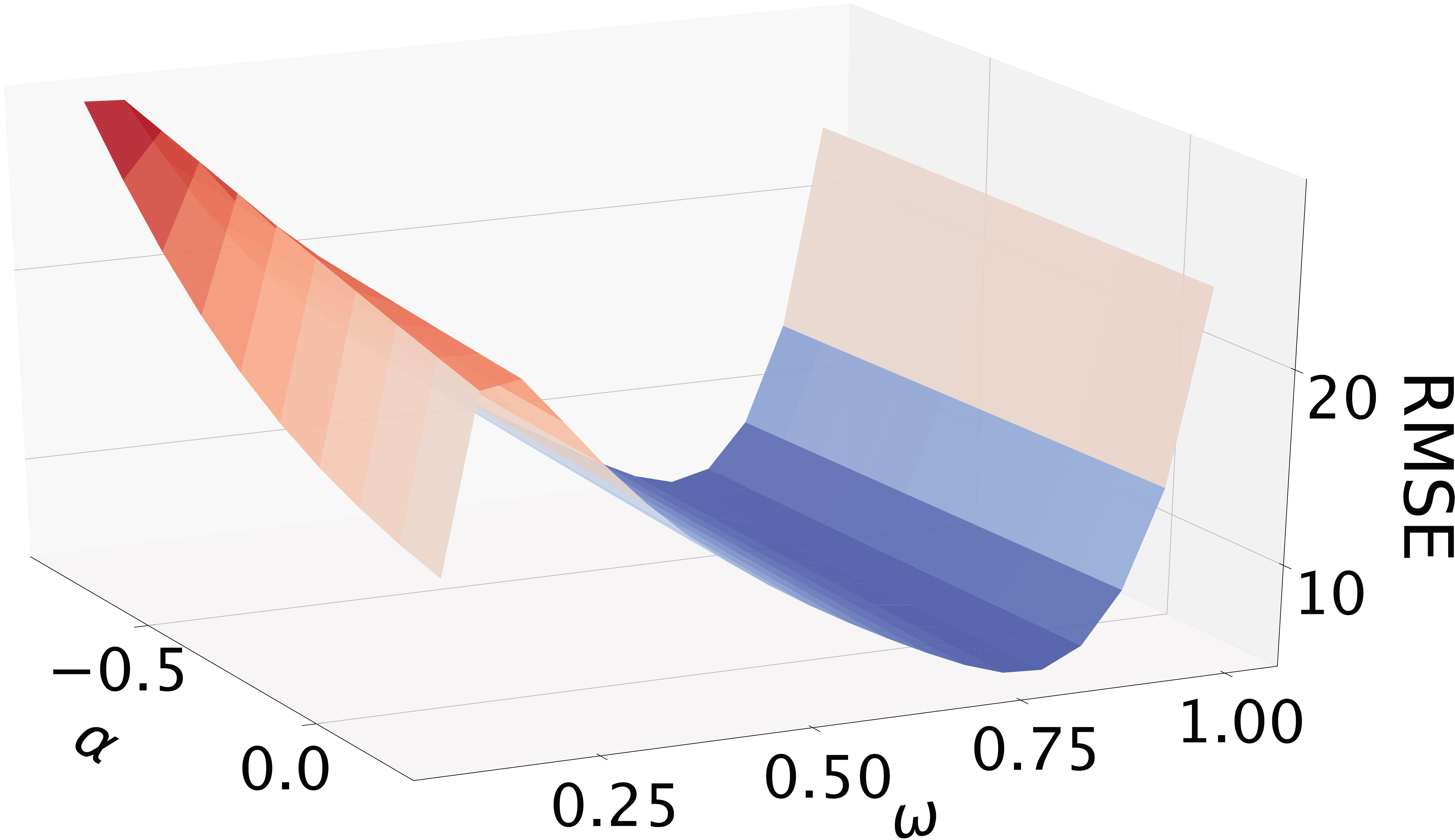}
    \includegraphics[width=.3\linewidth]{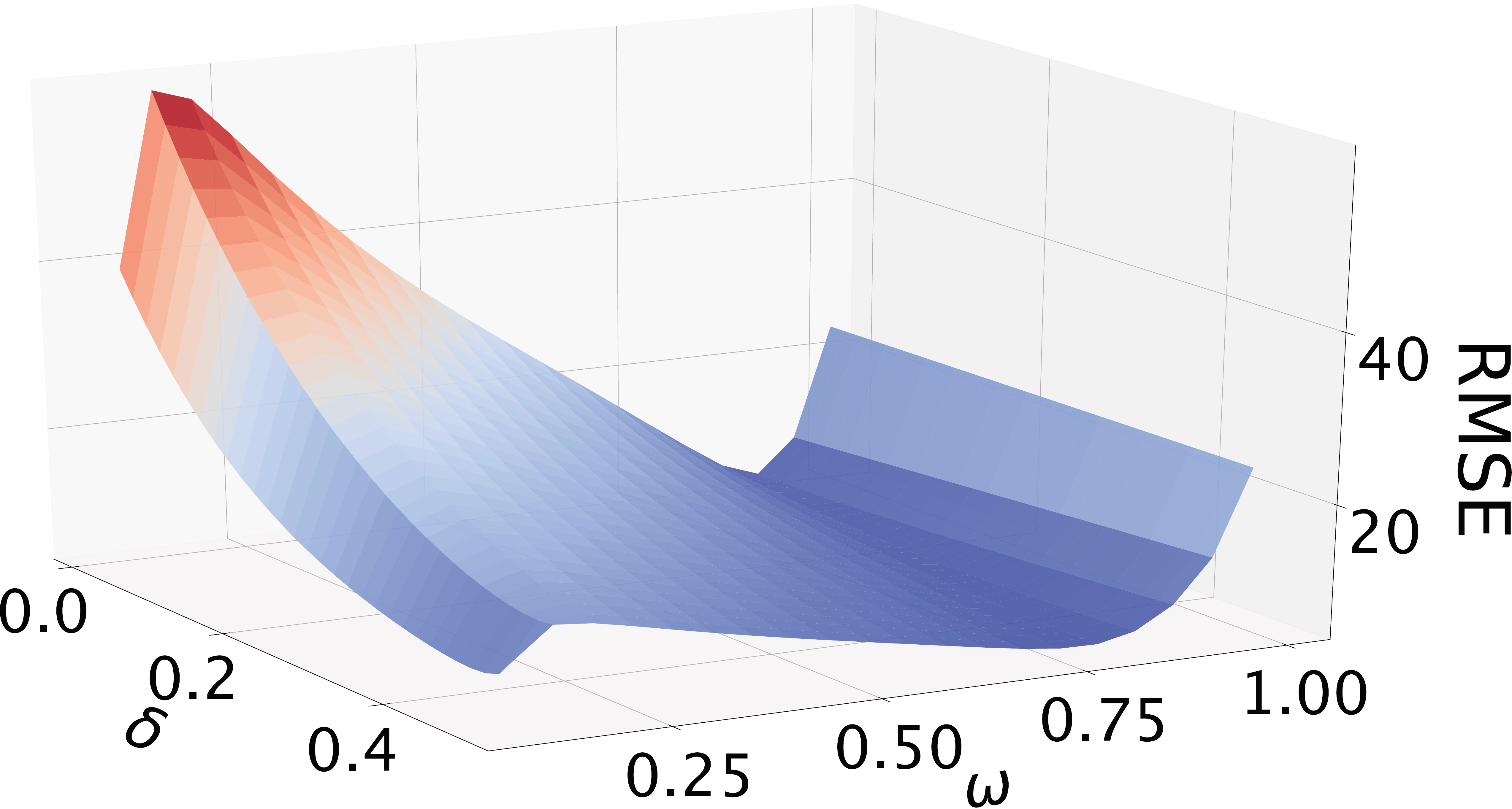}
    \caption{
    	\emph{left}: predictions made using one week's worth of history;
        \emph{center}: analysis of $\alpha$ and $\omega$
            for $\delta = \sqrt{\omega}/(1+\sqrt{\omega})$;
        \emph{right}: analysis of $\delta$ and $\omega$
            for fixed $\alpha=0$.
        For $\omega$ too small, ill-conditioning plays a 
        confounding role that requires further consideration.
        \label{fig:cali_results}
    }
\end{figure}

We explore the implications of varying the free parameters
$\alpha$, $\omega$ and $\delta$ in \figref{fig:cali_results}.
To simplify the analysis, the Gaussian kernel component has a
single fixed length scale in both dimensions of $.01$.
We also use a fixed noise variance term (Tikhonov regularization
parameter) of $10^{-8}$, primarily to avoid ill-conditioning.
In principle, these quantities should be determined from
the data.


\figref{fig:cali_results} suggests that for many values of
$\omega$ the predictive power is unaffected by a wide range of
$\alpha$ values.  Similarly, though to a lesser extent, $\delta$
appears to have a limited and somewhat predictable effect on 
performance provided $\omega$ is reasonably large.

\section{Future Work}
This initial analysis suggests that there is an opportunity for
this new kernel to help in fitting and predicting space-time
data.  The free parameters provide interpretable flexibility in
modeling, but this comes at the cost of choosing them
appropriately.  Doing so will be a fundamental component of
future research on the viability of this tool.

Of particular interest will be
understanding how the nonstationarity and asymmetry
of the kernel manifests itself in predictions at future times.
Initial experimentation, including some omitted for space 
considerations, suggests a nontrivial interplay
between $\delta$ and $\omega$;
furthermore, this relationship seems to vary as observed data
(and, thus, the desired region of prediction) advances from $s=0$.
One future goal is to identify how existing prediction strategies,
including cross-validation and maximum likelihood estimation,
can be effectively adapted for this kernel.




\bibliographystyle{plain}

\end{document}